\documentclass[12pt,a4paper,final]{iopart}

\usepackage{graphicx}
\usepackage{graphics}
\usepackage{amssymb}
\usepackage{amsmath}
\usepackage{epsfig}
\usepackage{latexsym}
\usepackage{color}
\usepackage{rotating}
\usepackage{subfigure}
\usepackage{hyperref}
\usepackage{times}

\newcommand{\beq}{\begin{equation}}
\newcommand{\eeq}{\end{equation}}
\newcommand{\bea}{\begin{eqnarray}}
\newcommand{\eea}{\end{eqnarray}}
\newcommand{\bean}{\begin{eqnarray*}}
\newcommand{\eean}{\end{eqnarray*}}
\makeatother

\begin{document}

\title{Shaking photons from the vacuum: acceleration radiation from vibrating atoms}

\author{Brian P. Dolan}
\address{Department of Theoretical Physics, Maynooth University, 
  Maynooth, Co.~Kildare, Ireland}
\address{School of Theoretical Physics,
  Dublin Institute for Advanced Studies,
  10 Burlington Rd., Dublin, Ireland}

\author{Aonghus Hunter-McCabe}
\address{Department of Theoretical Physics, Maynooth University,
  Maynooth, Co.~Kildare, Ireland}

\author{Jason Twamley}
\address{Centre for Engineered Quantum Systems, Department of Physics and Astronomy,
Macquarie University, Sydney, New South Wales 2109, Australia}

\begin{abstract}
Acceleration radiation - or Unruh radiation -  the thermal radiation observed by an ever accelerating observer or detector, although having similarities to  Hawking radiation, so far has proved extremely challenging to observe experimentally. One recent suggestion is
that, in the presence of a mirror,  constant acceleration of an atom in its ground state can excite 
the atom while at the same time cause it to emit a photon 
in an Unruh-type process. In this work we show that merely by shaking the atom, in simple harmonic motion for example,
can have the same effect.
We calculate the transition rate for this in first order perturbation theory and consider harmonic motion of the atom in the presence of a stationary mirror, or within a cavity or just in empty vacuum. For the latter we propose a circuit-QED potential implementation that yields transition rates of $\sim 10^{-4}\,{\rm Hz}$, which may be detectable experimentally.
\end{abstract}

\maketitle
\section*{Introduction}
The study of quantum fields in curved space time has led to profound new discoveries including Hawking radiation and the Unruh \cite{Unruh1976} effect. The latter typically is discussed for the case of an uniformly accelerated observer or detector in the vacuum of flat Minkowski spacetime. The accelerated detector (which we take to be a two level system - TLS), rather than seeing the vacuum, instead experiences a thermal photon bath whose temperature $T=a\hbar/(2\pi c k_B)$, where $a$ is the proper acceleration of the TLS. One interpretation is that the virtual photons that normally dress  the internal states of such a TLS are promoted to be real excitations due to the highly non-adiabatic nature of the acceleration. Virtual photons can have measurable signatures in atomic physics, e.g. in the Lamb shift and in Raman scattering. However, experimentally observing the Unruh effect has proved challenging since to achieve $T\sim 1{}^\circ$ K, requires an extreme acceleration $a\sim 10^{20} \,{\rm m/s^2}$, and to substantially excite the TLS  the latter should have a transition frequency $\omega_0/2\pi\sim 20\,{\rm GHz}$.
The importance of the Unruh effect and its analogous effect in black holes, Hawking radiation, has led to a number of proposals over the past three decades towards an experimental test of the existence of acceleration radiation.  These proposals include detecting Unruh radiation  via electrons orbiting in storage rings \cite{Bell1983, Bell1987, Unruh1998},  in Penning traps \cite{Rogers1988}, in high atomic number nuclei \cite{Rad2012},  via shifts in accelerating Hydrogen-like atoms \cite{Pasante1998}, via   decay processes of accelerating protons or neutrons \cite{Matsas1999}, or when electrons experience ultra-intense laser acceleration \cite{Chen1999,Schutzhold2006}, by examining the Casimir-Polder coupling to an infinite plane from an accelerating two-level system \cite{Rizzuto2007}. Researchers have also investigated using cavities to enhance the effect \cite{Scully2003, Belyanin2006,Lopp2018}, and using the Berry phase or entanglement as probes of Unruh radiation \cite{Martin-Martinez2011,Hu2012,Tian2017}. With the advent of circuit QED - cQED, researchers have investigated simulations of Unruh radiation via the Dynamic Casimir Effect - DCE, \cite{Johansson2009,Wison2011, delRey2012, Lahteenmaki2013}, or by using cQED to simulate relativistically moving systems \cite{Felicetti2015,Su2017}, and also using NMR \cite{Jin2016} or by studying the interaction between pairs of accelerated atoms \cite{Rizzuto2016}, or via the dynamical Casimir effect \cite{Farias2019}. More recent work has probed  whether real motion can produce acceleration radiation and in \cite{Sanz2018,Wang2019}, the authors consider a mechanical method of modulating the electromagnetic fields in cQED DCE photon production.

\begin{figure}[tp] 
\begin{center}
\setlength{\unitlength}{1cm}
\begin{picture}(8.5,7)
\put(-2,-.2){\includegraphics[width=.85\columnwidth]{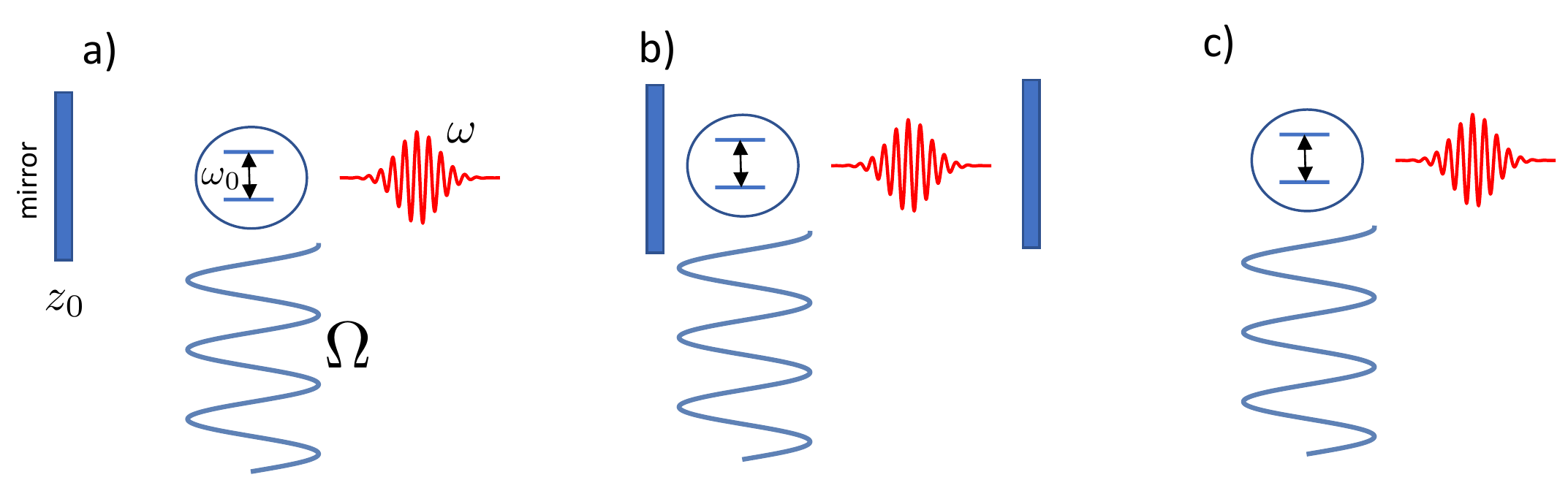}}
\end{picture}
\end{center}
\caption{We consider the generation of photons of frequency $\omega$, from a two level system (atom), with internal transition frequency $\omega_0$, mechanically oscillating at frequency $\Omega$ and amplitude $A$, and initially in the ground state with the electromagnetic field in the vacuum. a) oscillating in front of a mirror, b) oscillating inside a cavity, c) oscillating in free space.}
   \label{Fig1}
\end{figure}

In our work we instead consider a model where the centre of mass of a TLS moves in an accelerated manner   that may  be achieved in a laboratory setting e.g. oscillatory motion.  In \cite{Svidzinsky2018} the authors discussed the possibility that a TLS, uniformly accelerating away from a mirror and initially in its ground state, could experience a transition to its excited state
accompanied by the emission of a  photon.   This raises the question of what other kinds of acceleration might lead to such a process? In this paper we show that simple harmonic oscillation of the atom can also result in photon emission accompanied by an excitation from the ground  to the excited state of the TLS. 

{\color{black} Considering acceleration radiation from oscillatory motion has an advantage over continuous linear accelerated motion in that the TLS stays in a compact region and thus should be more feasible for direct experimental implementation.} In the following we derive closed compact expressions for the rate of photon production in the case of an oscillating TLS in the presence of a mirror, or within cavity, or just coupled to vacuum. For the latter we propose a cQED implementation which predicts significant rates of photon production in the microwave regime.
\vspace{.25cm}\\

{\color{black}\section{\it\bf Evaluating the probability to emit a photon under general non-relativistic motions:-} We now show how to derive the probability for the two-level system, moving in one dimension along a pre-set space-time trajectory, in the presence of a mirror, to become excited and emit a photon using first order perturbation theory.  Using the derived expression we first confirm that if the two level system is at rest the probability to emit a photon is zero. In the subsequent subsections we consider various other types of motions.} Following \cite{Svidzinsky2018}, we consider a two level system, or atom, coupled to the  electromagnetic field with a coupling strength $g$, and atomic transition frequency  $\omega_0$, so the excitation energy is $\Delta E=\hbar \omega_0>0$. 
For simplicity we consider motion in
one-dimension given by position $z(t)$ 
\footnote{Unlike \cite{Svidzinsky2018} our analysis will be non-relativistic and $t$ is just ordinary Newtonian time, not proper time --- the calculation can be done relativistically but there is no need for such a complication. If we assume a mechanical oscillation frequency $\Omega/2\pi \sim\, 10\, {\rm GHz}$, and a maximum amplitude of oscillation as $A\sim 10 {\rm nm}$, then the maximum velocity achieved of the atom is $v_{max}\sim 600\, {\rm m/s}\ll c$. As we show below we predict that these parameter values will yield significant acceleration radiation.}  
In the interaction picture the Hamiltonian for the interaction of the atom with the electric field
 $\phi_\omega$, is \cite{Svidzinsky2018},
\begin{eqnarray}H_I(\omega,t) = \hbar g\bigl\{a_\omega \phi_\omega(t,z(t))
+ a^\dagger_\omega \phi^*_\omega(t,z(t)) \bigr\}\\\label{noSMA}
\times \bigl(\sigma_- e^{-i\omega_0 t} + \sigma_+ e^{i\omega_0 t}  \bigr)\label{Hamomega}
\end{eqnarray}
where $\sigma_+$ raises the internal atomic state and $\sigma_-$ lowers it, $a_\omega^\dagger$, $a_\omega$ are photon creation operator and annihilation operators and
$\phi_\omega$ are field modes that depend on the boundary conditions. {\color{black} We note that (\ref{Hamomega}), describes the interaction at a specific frequency $\omega$, and the full Hamiltonian is obtained by $H_I(t)\equiv \int\,H_I(\omega, t)\,d\omega$, and thus our treatment encapsulates the potential excitation of any wavelength radiation. In what follows we compute the probability to excite the atom and emit a photon of frequency $\omega$, where the latter is not taken as a fixed quantity.}
{\color{black} This interaction has been used numerous times in the literature to model the coupling between a two-level atom and a quantum field but here we do not make the rotating wave approximation and  the position of the atom is allowed to vary in time. Further, we permit the photon field mode frequency $\omega$ to be arbitrary and thus the the atom can couple to vacuum modes of any frequency. This is unlike the so-called ``single mode approximation''  (SMA), where authors consider the atom to couple preferentially to a small number of modes concentrated at a single frequency e.g. for an atom within an accelerating cavity \cite{Alsing2003, FSchuller2005, Alsing2006, Datta2009, Wang2010, Dragan2011, Hwang2012, Brown2012, Lopp2018}. Research has shown that making the SMA can lead to difficulties in superluminal propagation effects at strong couplings \cite{Sanchez2018}, and entanglement generation \cite{Bruschi2010}, however some works have  explored beyond the SMA including an NMR analogue simulation of Unruh radiation \cite{Jin2016} [2 modes].    }
If the two level system is moving on the entire real line and its position is a function of $t$, we would use right and left moving field modes
\bea \phi_\omega=e^{-i (\omega t \mp  k z(t))}, \label{eq:infinite-phi}  \eea
where $k=\omega/c>0$, is the $z$-component of the photon's wave-vector.
In the presence of a mirror fixed at $z=z_0$, we would instead use
\beq \phi_\omega=e^{-i(\omega t - k z(t) + k z_0)}- e^{-i (\omega t + k z(t) - k z_0)},\label{eq:mirror-phi}\eeq
thus ensuring that the photon field, and hence the transition amplitude,  vanishes when $z(t)$ is an integer multiple of the wavelength and the atom is at a node of the field.  In a cavity of length $L$, we would again use 
(\ref{eq:mirror-phi}), but the frequency and wave-vector would be restricted by the condition that  $k=\omega/c=2 \pi   n/L$ for some positive integer $n$.
We shall work with (\ref{eq:mirror-phi}) on the half-line for the moment,
and later adapt  the results to the case of a cavity or the entire real line as in (\ref{eq:infinite-phi}). 
In first order perturbation theory the probability of exciting the atom (via the raising operator $\sigma_+ e^{i\omega_0 t}$ in the interaction), and at the same time creating  a photon {\color{black}  of frequency $\omega$ }
via the $\hat a^\dagger \phi_\omega^*$,
term in the interaction potential, is given by 
\beq P= g^2 \left|
\int_{-\infty}^\infty\bigl[
e^{i k (z(t)- z_0)} -c.c. \bigr] e^{i \omega t + i \omega_0 t}
d t \right|^2 .\label{eq:phi-mirror}\eeq
{\color{black} If the atom is at rest, $z$ is constant and  the integral gives $\delta(\omega + \omega_0)$, which is zero since $\omega>0$ and $\omega_0>0$.  Thus $P=0$, and the probability to become excited and emit a photon vanishes. We now use the expression (\ref{eq:phi-mirror}), in the following to consider other types of space-time motions $z(t)$, to discover how they can give rise to non-vanishing probabilities $P$.}
\vspace{.25cm}\\

\section{\it\bf Oscillating 2-state atom with a mirror :-}\label{mirror}
We now consider an atom forced to oscillate in the presence of a stationary mirror, the latter located at $z_0$, oscillating with a motional angular frequency $\Omega$,  around $z=0$
with amplitude $0<A<z_0$ 
(so we do not hit the mirror),  and set $ z(t) = A \sin(\Omega t).$
Using this in (2), the probability of creating a photon of frequency
$\omega$ from the vacuum, and at the same time exciting the atom from its ground state to its excited state is
$$P(\omega)= g^2 \left|
\int_{-\infty}^\infty\bigl[
e^{i k (A \sin(\Omega t) - z_0)} -c.c. \bigr] e^{i (\omega +  \omega_0)t} d t
\right|^2. $$
Simplify the notation by absorbing $\Omega$ into $t$ and defining dimensionless variables 
$\tau = \Omega t$, $\tilde \omega = (\omega +\omega_0)/\Omega$,
$\tilde A = k A$ and $\tilde z_0 = k z_0$, we obtain
\begin{eqnarray} P(\omega) &=& \frac{g^2}{\Omega^2} \left|
\int_{-\infty}^\infty\bigl[
e^{i (\tilde A \sin \tau - \tilde z_0)} -c.c. \bigr] e^{i \tilde \omega \tau} d \tau\right|^2 \nonumber\\ 
&=& \frac{g ^2}{\Omega^2} \left|
e^{-i\tilde z_0}\int_{-\infty}^\infty e^{i (\tilde A \sin \tau+ \tilde \omega \tau)} d \tau \right.\nonumber\\ 
& &\qquad\qquad\left.-e^{i\tilde z_0}\int_{-\infty}^\infty  e^{i (\tilde A \sin \tau - \tilde \omega \tau)} d  \tau\right|^2.
\label{totalSHO-probability} 
\end{eqnarray}

The integrals appearing in (\ref{totalSHO-probability}) 
 are related to
Anger functions \cite{Abramowitz1970} (section 12.3.1),
$$ {\bf J}_\nu(x)=\frac{1}{2\pi}\int_{-\pi}^\pi e ^{i(x \sin\theta - \nu \theta)}d \theta
= \frac{1}{\pi}\int_0^\pi \cos(x\sin\theta - \nu\theta)d \theta.$$
But Anger functions are not quite what we want since what we have in (\ref{totalSHO-probability})
, is
\begin{eqnarray}
\int_{-\infty}^\infty  e ^{i(x \sin\theta - \nu \theta)} d \theta
&=&\!\! \sum_{s=-\infty}^\infty\, \int_{-\pi}^\pi e^{i\bigl(x \sin\theta - \nu(\theta+2s\pi)\bigr)}d \theta \\
&=&\!\! \pi {\bf J}_\nu(x) \sum_{s=-\infty}^\infty  e^{-2is\nu\pi}.
\label{eq:totalamplitude}
\end{eqnarray}
Letting $ \nu = n + \{\nu\}$, 
where $n$ is an integer, and $0 \le  \{\nu\}<1$, is the non-integral part of $\nu$,
then
$ \sum_{s=-\infty}^\infty  e^{-2is\nu\pi}= \sum_{s=-\infty}^\infty  e^{-2is\{\nu\}\pi} = 1 + 2\sum_{s=0}^\infty  \cos(2s\{\nu\}\pi) = \pi \delta(\{\nu\})$
and (\ref{eq:totalamplitude}), vanishes unless $\nu$ is an integer, in which case it
diverges. This is not unexpected since when Fermi's Golden Rule says that if the transition rate is constant, integrating over all $t$ will necessarily give an infinite answer for any transition with non-zero probability.  For our periodic case its more informative to estimate a transition rate over a motional cycle rather than the accumulated  probability over all time.

\sloppy Before estimating this let us consider the case when $\nu$ is a non-zero rational number, $\nu = p/q$ with $p$ and $q$ mutually prime and positive, in the integral
$$ \int_{-\infty}^\infty e^{i(x\sin \theta -\nu \theta)} d \theta
= q \int_{-\infty}^\infty e^{i(x\sin (q \psi)  - p \psi)} d \psi 
$$
where $\psi = \theta/q$. The integrand is periodic in $\psi$
with period $2\pi$ so again the integral will diverge (unless the integral over one period vanishes).  Integrating over just one period in $\psi$, we can define
\begin{eqnarray}
{\cal J}(x;p,q) &=& 
\frac{1}{2 \pi} \int_{-\pi}^\pi e^{i(x \sin (q \psi)  - p \psi )} d \psi\\\nonumber
&=& \frac{1}{\pi} \int_0^\pi \cos\big\{x \sin (q \psi)  - p \psi \bigr\} d \psi,
\label{eq:J-def}
\end{eqnarray}
which, in terms of Anger functions, becomes
\begin{eqnarray}
{\cal J}(x;p,q) &=& 
\frac{1}{2 \pi q} \int_{-q\pi}^{q \pi} e^{i(x \sin \theta  - \frac p q  \theta )} d \theta \\\nonumber
&=&\frac{1}{q}  {\bf J}_{\frac p q }(x)\sum_{s=0}^{q-1} e^{i \pi (2 s - q + 1)\frac{p}{q}}.\nonumber
\end{eqnarray}
We note however that $$\sum_{s=0}^{q-1} \exp({i \pi (2 s - q + 1)p/q}) = 
\exp({-i\pi\frac{p(q-1)}{q}}) \sum_{s=0}^{q-1} \exp({2i \pi p s/q})=0,$$ 
since $\sum_{s=0}^{q-1} \exp(2i \pi p s/q)$,
vanishes for any two integers $p$ and $q$, provided $p/q$ is not an 
integer.
 From this observation we conclude  that the transition rate is zero unless $\tilde \omega$ is a
positive integer, that is 
\begin{equation}
\omega + \omega_0 = n\Omega,\label{eq:Omega-restriction}
\end{equation}
where $\Omega$ is the mechanical frequency, $\omega_0$ is the atomic transition frequency, $\omega$ is the frequency of the photon field and $n$ is an integer.
Since $\tilde \omega$ is  positive, $n>0$, and to obtain a closed expression for the transition probability over a mechanical cycle we can use the integral
representation of the Bessel function,  \cite{Abramowitz1970} (section 9.1.21),
$J_n (\tilde A) = \frac{1}{2 \pi}\int_{-\pi}^{\pi} 
\exp[{i(\tilde A \sin \tau - n \tau)}] d \tau = (-1)^n J_{-n}(\tilde A)$,
to write  $\int_{-\pi}^{\pi} \exp [{i(\tilde A \sin \tau + \tilde z_0 -  n  \tau)} ]d \tau= 2 \pi e^{i \tilde z_0}  J_n(\tilde A)$, and $\int_{-\pi}^{\pi} \exp[{i(\tilde A \sin \tau -
 \tilde z_0 +  n \tau)} ]d  \tau
= 2 \pi e^{-i (\tilde z_0 - n \pi)}  J_n(\tilde A)$.
By dividing by the period of mechanical oscillation, $2\pi/\Omega$, we get a transition rate (in ${\rm Hz}$), as
\begin{eqnarray}
\overline P_{n} & = &\frac{\Omega }{2 \pi}
\frac{g^2}{ \Omega^2} \left|
 \Bigl[ 2 \pi \left(e^{-i (\tilde z_0 - n \pi)} -  e^{i \tilde z_0}\right) \Bigr]
J_n(\tilde A)\right|^2\nonumber\\
&=&
\frac{8 \pi g^2}{\Omega} \sin^2\left(\tilde z_0 -\frac{\pi n}{2}\right) J_n^2(\tilde A)\\
&=&
\frac{8 \pi g^2}{\Omega} \sin^2\left(k z_0 -\frac{\pi n}{2}\right) J_n^2(k A),\label{eq:mirror-rate}
\end{eqnarray}
where $\omega = n \Omega - \omega_0 >0$. 
\vspace{.25cm}\\

\section{\it\bf 2-dimensional motion}

The formalism can be applied to a 2-level atom following any closed trajectory $\vec x(t)$ in the two dimensional $y-z$ plane, with a flat mirror located at $z_0$ extending in the $y$-direction.   For an electromagnetic wave with wave-vector $(k_y,k_z)$ we simply replace $k (z(t) - z_0)$ in (\ref{eq:phi-mirror}) with
$\vec k.\vec x(t) - k_z z_0$. One cannot obtain analytic answers for a general trajectory but some simple cases are immediate:
\vspace{.25cm}\\

{\it\bf 2-level atom oscillating parallel to a mirror:-}  for oscillation in the $y$-direction replace $k (z(t) - z_0)$ in (\ref{eq:phi-mirror}) with
$ k_y A \sin(\Omega t)  - k_z z_0 $  with $A$ again a constant amplitude.
The analysis is identical and (\ref{eq:mirror-rate}) still holds, but with $k z_0$ replaced with $k_z z_0$ and $k A$ replaced with $k_y A$.
\vspace{.25cm}\\

{\it\bf Rotating 2-level atom with a mirror:-} for an atom rotating around the fixed point $(0,z_0)$ in a circle with radius $R$
and constant angular velocity $\Omega$, $\vec x(t) = R(\cos\Omega t),\sin(\Omega t)$ and
$\vec k.\vec x(t)=k_y R \cos(\Omega t) + k_z R \sin(\Omega t)$.
If we parameterize the direction of the electromagnetic wave by a phase $\delta$ with
 $k_y=k\sin\delta, \quad k_z = k\cos\delta$
then
  $k_y R \cos(\Omega t) + k_z R \sin(\Omega t) - k_z z_0  = k R \sin(\Omega t + \delta) - k z_0 \cos \delta $
and the phase $\delta$ 
in the sine function can be
absorbed into $\tau$ in (\ref{eq:phi-mirror}), which does not affect the result. We just replace the
amplitude $A$ in (\ref{eq:mirror-rate}) with the rotation radius $R$ and replace 
$k z_0$ with $k_z z_0 \cos \delta$.
\vspace{.25cm}\\

\section{\it\bf Oscillating 2-level atom in a cavity:-}\label{cavity}
 In a one-dimensional cavity containing no photons the transition rate to excite the atom and at the same time emit a photon of frequency $\omega$ is still given by (\ref{eq:mirror-rate}), except that the allowed values of $\omega$ are discrete, 
 $\omega = c k = \pi m c/L$,
with $m$ a positive integer and Eqn (\ref{eq:Omega-restriction}) imposes the condition
\[ \frac{\pi m c}{L} + \omega_0=n \Omega\]
on $\Omega$. The rate is enhanced however if the cavity already contains $N$ photons of 
frequency $\omega$, since then $\langle N+1|a^\dagger_\omega|N\rangle = \sqrt{N+1}$,
giving the transition rate to excite/de-excite the atom given $N$ photons in the cavity as,
\begin{equation}
\overline P_{n,m,N,\pm} = 
\frac{8 \pi \chi_\pm\, g^2}{\Omega} \sin^2\left(\frac{\pi m z_0}{L} -\frac{\pi n}{2}\right) J_n^2\left(\frac{\pi m A}{L}\right),\label{eq:cavity-rate}
\end{equation}
where $\chi_+=N+1$, to excite and emit a photon, while $\chi_-=N$ to de-excite the atom and absorb a photon, and
$n \Omega = \omega_0 - \omega = \omega_0 - \pi m c/L$. {\color{black} We note that in order to observe this phenomena one could consider one of the mirror's to be slightly imperfect and then one could perform spectroscopy on the photons leaking from the optical cavity. The signature of the photons created via (\ref{eq:cavity-rate}), will prove challenging to detect over any background of existing photon occupation within the cavity.}
\vspace{.25cm}\\

\begin{figure}[tp] 
\begin{center}
\setlength{\unitlength}{1cm}
\begin{picture}(8.5,7.5)
\put(-2,-.3){\includegraphics[width=.65\columnwidth]{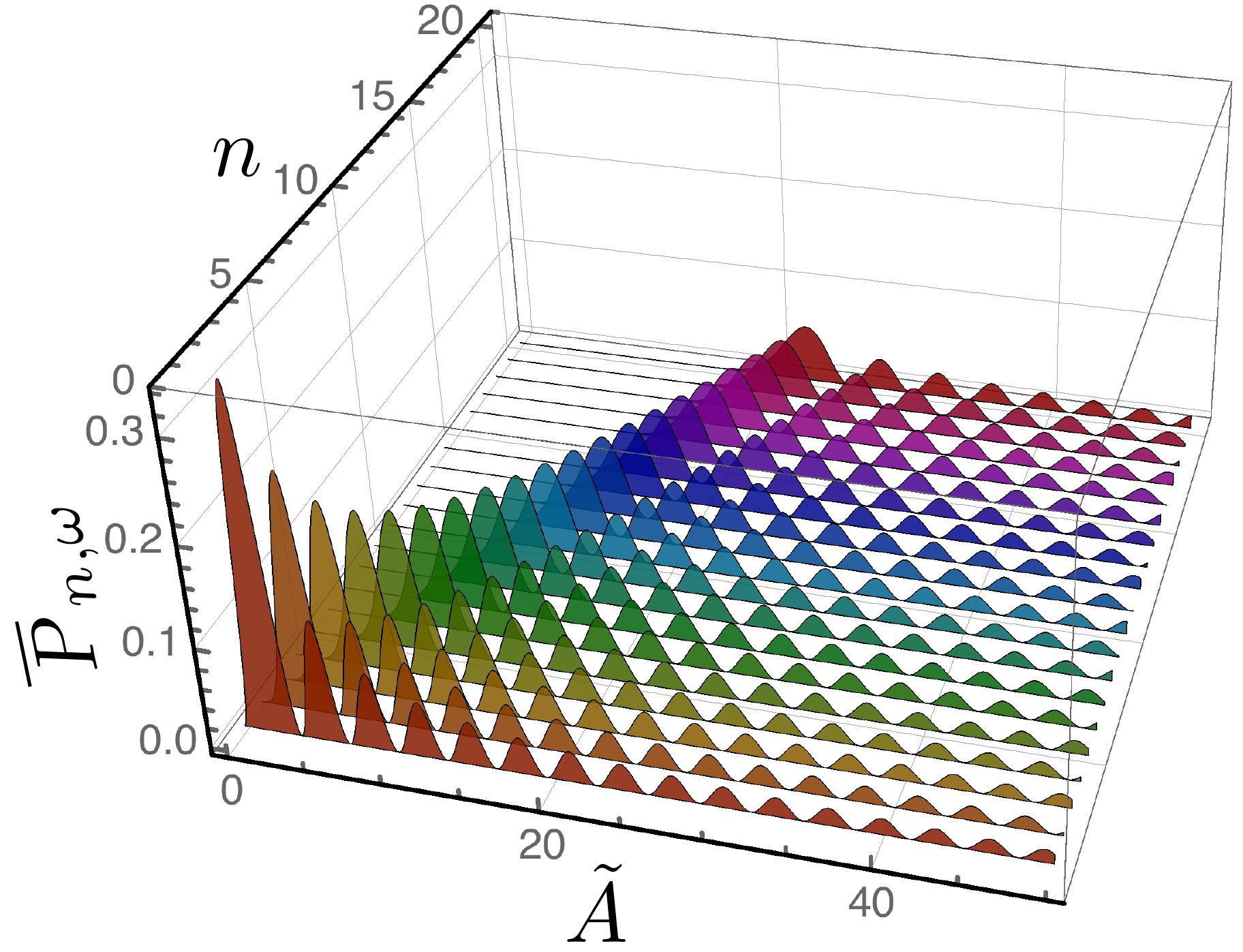}}
\end{picture}
\end{center}
\caption{Transition rate for SHO forced atom in free space to emit a photon with angular frequency $\omega = n\Omega - \omega_0$ and wavelength $\lambda={2\pi c}/{\omega}$, equation  (\ref{eq:P_n-no-mirror}), as a function of dimensionless oscillation amplitude $\tilde A =({2\pi}/{\lambda}) A$ and sideband index $n$. The prefactor 
${2 \pi g^2}/{\Omega}$ is omitted.  We see that transition rate is negligible until $\tilde A$ is of order $n$.}
   \label{Fig2}
\end{figure}

\section{\it\bf Atom performing SHO in free space:-}\label{free} 
Repeating the analysis of previous section when there's no mirror, the transition rate to emit a   photon is,  using (\ref{eq:infinite-phi}),
\begin{equation}
\overline P_{n,\omega}  = \frac{\Omega }{2 \pi}
\frac{g^2}{\Omega^2} \left|
\int_{-\pi}^\pi e^{i \tilde A \sin \tau} e^{i \tilde \omega \tau} d \tau
\right|^2 
  = \frac{2 \pi g^2 }{\Omega} J_n^2( k A) ,
\label{eq:P_n-no-mirror}
\end{equation}
with $n>0$. Even with no mirror an atom can be excited and emit a photon at the same time by shaking the atom hard enough to supply the required energy.  
The transition rate (\ref{eq:P_n-no-mirror}), with  $\omega = n\Omega - \omega_0$ is 
\begin{equation}
\overline P_{n}
  = \frac{2 \pi g^2 }{\Omega} J_n^2\left(\frac { (n \Omega -\omega_0)  A }{c}\right),
\label{SHOfreespace}  
\end{equation}
and for a given $n$ the rate will be largest at the first peak of $J_n$. From Fig \ref{Fig2}, we observe that for a given $n$, the transition rate is negligible unless $\tilde{A}\sim n$. The maximum rate occurs when $n=1$ and $\tilde{A}\sim 1.8$, and here $\overline P_1\approx 2.1g^2/\Omega$, but to achieve this one must have a mechanical oscillation amplitude of $A\sim 1.8 c/(\Omega-\omega_0)$, which will be extremely large when compared to the highest realistically achievable values of $\Omega-\omega_0$, in an experiment. 

\begin{figure}[tp] 
\begin{center}
\setlength{\unitlength}{1cm}
\begin{picture}(8.5,7.5)
\put(-2,-.3){\includegraphics[width=.65\columnwidth]{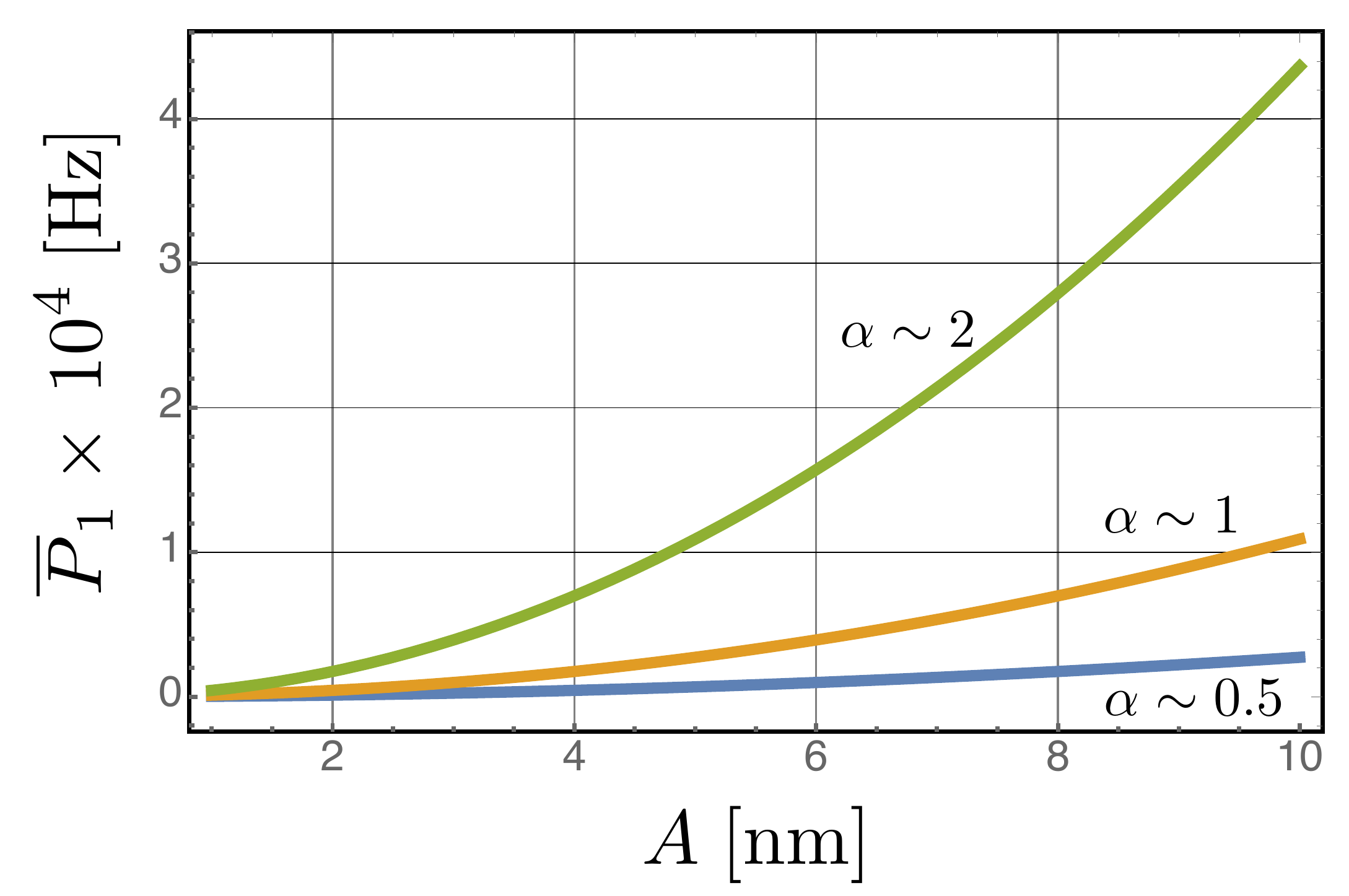}}
\end{picture}
\end{center}
\caption{Transition/emission rate for an atom oscillating with simple harmonic motion in free space as a function of the amplitude of the oscillation  and ultra-strong coupling strengths $\alpha$, where $g=\alpha \omega_0$, taking a mechanical frequency $\Omega/2\pi=10\,{\rm GHz}$, and two level transition frequency $\omega_0=\Omega/2$ (\ref{SHOfreespace}).}
   \label{Fig3}
\end{figure}

For a possible experimental demonstration of acceleration from a two level atom shaking in a vacuum we instead consider the case when the amplitude of oscillation $A$ is small enough so that the argument of the Bessel function in (\ref{SHOfreespace}), is small and we can use the expansion $J_1^2(x)\sim x^2/4,\; 0<x\ll 1$. To achieve a large rate we consider ultra-strong coupling $g$ of the atom to the continuum where $g=\alpha \omega_0$, with $\alpha\sim 0.2$ and above \cite{Forn-Diaz2016,  Magazzu2018, PuertasMartinez2019, FriskKockum2019, Forn-Diaz2019}. We consider a cQED setup where a two level superconducting artificial atom is placed on a vibrating cantilever/membrane which is ultra-strongly coupled via an inductive coupling to the electromagnetic vacuum of a nearby electromagnetic continuum of a superconducting co-planar waveguide. This is similar to recent experimental work \cite{Viennot2018}. By using a diamond based  nanomechanical resonator one can achieve motional frequencies in the GHz regimes \cite{Gaidarzhy2007,Rodriguez-Madrid2012}. From (\ref{SHOfreespace}),  with the approximation $J_1(x)^2 \sim x^2/4$, the rate is maximised when $\omega_0=\Omega/2$, and taking this for small amplitude oscillations we obtain $\overline P_{1}\approx(\pi A\alpha)^2\Omega^3/(32 c^2)$. Fig \ref{Fig3}, shows that the emission/transition rate for a shaken superconducting two level system can reach values of $\overline P_1\sim 10^{-4}\,{\rm Hz}$, which may be detectable in current cQED setups. {\color{black} One might also consider placing the mechanically oscillating qubit close to a mirror as discussed in Section \ref{mirror}. Oscillatory motion along an arbitrary direction in three dimensions would couple the qubit to both vacuum modes which either have the mirror as a boundary or unbounded vacuum modes (for motion parallel to the mirror). This would result in  emission of photons perpendicular to the mirror (the case treated in Section \ref{mirror}), and in directions parallel to the mirror's plane (the case treated in Section \ref{free}). Such a situation may be difficult to observe experimentally as one must achieve large couplings between the two level system and the continuum. This is possible in the above discussed cQED setup above as the electromagnetic continuum is restricted essentially to be one dimensional.}

{\color{black} We finally compare our results with recent related works. In \cite{Lo2018}, and \cite{Ferrari2019}, the authors respectively consider acceleration radiation from a mechanically oscillating two level system in free space  and the modification of spontaneous emission in a two level system adjacent to an oscillating mirror. However their studies are restricted to the case where $\Omega<\omega_0$, and only multi-photon off-resonant processes play a role in these cases. Such off-resonant processes cannot be captured via our first order perturbative analysis and thus we cannot make any clear comparison with their results. However the work of \cite{Souza2018}, which looks at the radiation emitted from a two-level atom oscillating in free-space, do consider the regime $\Omega>\omega_0$, and include resonant processes and find the  emission of photons with the frequency $\omega_1=\Omega-\omega_0$, but do not find the higher modes $\omega_n=n\Omega-\omega_0$, which we predict to also exist, though with greatly reduced probabilities. To make a more quantitative comparison we study the small photon frequency case when $\omega_1\equiv\Omega- \omega_0$, is small and $\omega_1 A/c\ll 1$.  Making the approximation $J_1(x)^2\sim x^2/4$, for $x\ll 1$, and comparing our rate (\ref{SHOfreespace}), with Eqn (5) \cite{Souza2018}, in this regime, (in the notation of \cite{Souza2018}, this is when $\omega_1\equiv\omega_{\rm cm}- \omega_0$ is small), we find that both rates scale as $\overline P_{1}\sim \Gamma_{\rm MIE}\sim A^2/\Omega$. However in our case we find $\overline P_{1}\sim \omega_1^2$, while in \cite{Souza2018}, $\Gamma_{\rm MIE}\sim \omega_1^3$, a difference which may be due to the differences between the Hamiltonians. In our study we assume the Hamiltonian (\ref{noSMA}),  a  model for coupling between two levels systems and vacuum fields used by many works and which also was used in \cite{Svidzinsky2018}, to derive the Unruh temperature for the case of uniform acceleration, while \cite{Souza2018}, Eqn (1), includes both the normal dipole coupling but also a term linear in the velocity of the two level system. }

\section*{Acknowledgments}
We acknowledge support from the Australian Research Council project CE170100009.

\vspace{.1cm}
\section*{References}

\begin{thebibliography}{10}
\expandafter\ifx\csname url\endcsname\relax
  \def\url#1{{\tt #1}}\fi
\expandafter\ifx\csname urlprefix\endcsname\relax\def\urlprefix{URL }\fi
\providecommand{\eprint}[2][]{\url{#2}}

\bibitem{Unruh1976}
Unruh W~G 1976 {\em Phys. Rev. D\/} {\bf 14} 870--892

\bibitem{Bell1983}
Bell J and Leinaas J 1983 {\em Nucl. Phys. B\/} {\bf 212} 131--150 ISSN
  05503213

\bibitem{Bell1987}
Bell J and Leinaas J 1987 {\em Nucl. Phys. B\/} {\bf 284} 488--508 ISSN
  05503213

\bibitem{Unruh1998}
Unruh W 1998 {\em Phys. Rep.\/} {\bf 307} 163--171 ISSN 03701573

\bibitem{Rogers1988}
Rogers J 1988 {\em Phys. Rev. Lett.\/} {\bf 61} 2113--2116

\bibitem{Rad2012}
Rad N and Singleton D 2012 {\em Eur. Phys. J. D\/} {\bf 66} 258

\bibitem{Pasante1998}
Passante ~R 1998 {\em Phys. Rev. A\/} {\bf 57} 1590  

\bibitem{Matsas1999}
Matsas G~E and Vanzella D~A 1999 {\em Phys. Rev. D\/} {\bf 59} 094004

\bibitem{Chen1999}
Chen P and Tajima T 1999 {\em Phys. Rev. Lett.\/} {\bf 83} 256--259 ISSN
  0031-9007

\bibitem{Schutzhold2006}
Sch{\"{u}}tzhold R, Schaller G and Habs D 2006 {\em Phys. Rev. Lett.\/} {\bf
  97} 121302
  
\bibitem{Rizzuto2007}
Rizzuto ~L  2007 {\em Phys. Rev. A\/} {\bf 76} 062411 



\bibitem{Scully2003}
Scully M~O, Kocharovsky V~V, Belyanin A, Fry E and Capasso F 2003 {\em Phys.
  Rev. Lett.\/} {\bf 91} 243004

\bibitem{Belyanin2006}
Belyanin A, Kocharovsky V~V, Capasso F, Fry E, Zubairy M~S and Scully M~O 2006
  {\em Phys. Rev. A\/} {\bf 74} 023807

\bibitem{Lopp2018}
Lopp R, Mart{\'{i}}n-Mart{\'{i}}nez E and Page D~N 2018 {\em Class. Quantum
  Gravity\/} {\bf 35} 224001 ISSN 0264-9381

\bibitem{Martin-Martinez2011}
Mart{\'{i}}n-Mart{\'{i}}nez E, Fuentes I and Mann R~B 2011 {\em Phys. Rev.
  Lett.\/} {\bf 107} 131301

\bibitem{Hu2012}
Hu J and Yu H 2012 {\em Phys. Rev. A\/} {\bf 85} 032105

\bibitem{Tian2017}
Tian Z, Wang J, Jing J and Dragan A 2017 {\em Ann. Phys. (N. Y).\/} {\bf 377}
  1--9

\bibitem{Johansson2009}
Johansson J~R, Johansson G, Wilson C~M and Nori F 2009 {\em Phys. Rev. Lett.\/}
  {\bf 103} 147003 ISSN 0031-9007

\bibitem{Wison2011}
Wilson C~M, Johansson G, Pourkabirian A, Simoen M, Johansson J~R, Duty T, Nori
  F and Delsing P 2011 {\em Nature\/} {\bf 479} 376--379

\bibitem{delRey2012}
del Rey M, Porras D and Mart{\'{i}}n-Mart{\'{i}}nez E 2012 {\em Phys. Rev. A\/}
  {\bf 85} 022511 ISSN 1050-2947

\bibitem{Lahteenmaki2013}
L{\"{a}}hteenm{\"{a}}ki P, Paraoanu G~S, Hassel J and Hakonen P~J 2013 {\em
  Proc. Natl. Acad. Sci.\/} {\bf 110} 4234--4238 ISSN 0027-8424

\bibitem{Felicetti2015}
Felicetti S, Sab{\'{i}}n C, Fuentes I, Lamata L, Romero G and Solano E 2015
  {\em Phys. Rev. B\/} {\bf 92} 064501

\bibitem{Su2017}
Su D, Ho C~T~M, Mann R~B and Ralph T~C 2017 {\em New J. Phys.\/} {\bf 19}
  063017 ISSN 1367-2630

\bibitem{Jin2016}
Jin F, Chen H, Rong X, Zhou H, Shi M, Zhang Q, Ju C, Cai Y, Luo S, Peng X and
  Du J 2016 {\em Sci. China Physics, Mech. Astron.\/} {\bf 59} 630302 ISSN
  1674-7348

\bibitem{Rizzuto2016}
 Rizzuto,~L,  Lattuca,~M,  Marino,~J\, {\it et al} 2016  {\em Phys. Rev. A\/} {\bf 94} 012121

\bibitem{Farias2019}
Far\'{\i}as, ~M ~B,  Fosco, ~C~D, Lombardo,~F~C and Mazzitelli,~F~D 2019 {\em Phys. Rev. D\/} {\bf 100} 036013

\bibitem{Sanz2018}
Sanz M, Wieczorek W, Gr{\"{o}}blacher S and Solano E 2018 {\em Quantum\/} {\bf
  2} 91

\bibitem{Wang2019}
Wang H, Blencowe M~P, Wilson C~M and Rimberg A~J 2019 {\em Phys. Rev. A\/} {\bf
  99} 53833

\bibitem{Svidzinsky2018}
Svidzinsky A~A, Ben-Benjamin J~S, Fulling S~A and Page D~N 2018 {\em Phys. Rev.
  Lett.\/} {\bf 121} 071301 ISSN 0031-9007

\bibitem{Alsing2003}
Alsing P~M and Milburn G~J 2003 {\em Phys. Rev. Lett.\/} {\bf 91} 1--4 ISSN
  10797114

\bibitem{FSchuller2005}
Fuentes-Schuller I and Mann R~B 2005 {\em Phys. Rev. Lett.\/} {\bf 95} 1--4
  ISSN 00319007

\bibitem{Alsing2006}
Alsing P~M, Fuentes-Schuller I, Mann R~B and Tessier T~E 2006 {\em Phys. Rev. A
  - At. Mol. Opt. Phys.\/} {\bf 74} 1--15 ISSN 10502947

\bibitem{Datta2009}
Datta A 2009 {\em Phys. Rev. A - At. Mol. Opt. Phys.\/} {\bf 80} 1--5 ISSN
  10502947

\bibitem{Wang2010}
Wang J, Deng J and Jing J 2010 {\em Phys. Rev. A - At. Mol. Opt. Phys.\/} {\bf
  81} ISSN 10502947

\bibitem{Dragan2011}
Dragan A, Fuentes I and Louko J 2011 {\em Phys. Rev. D - Part. Fields, Gravit.
  Cosmol.\/} {\bf 83} 1--4 ISSN 15507998

\bibitem{Hwang2012}
Hwang M~R, Jung E and Park D 2012 {\em Class. Quantum Gravity\/} {\bf 29} 0--9
  ISSN 02649381

\bibitem{Brown2012}
Brown E~G, Cormier K, Mart{\'{i}}n-Mart{\'{i}}nez E and Mann R~B 2012 {\em
  Phys. Rev. A - At. Mol. Opt. Phys.\/} {\bf 86} 1--9 ISSN 10502947

\bibitem{Sanchez2018}
{S{\'{a}}nchez Mu{\~{n}}oz} C, Nori F and {De Liberato} S 2018 {\em Nat.
  Commun.\/} {\bf 9} ISSN 20411723

\bibitem{Bruschi2010}
Bruschi D~E, Louko J, Mart{\'{i}}n-Mart{\'{i}}nez E, Dragan A and Fuentes I
  2010 {\em Phys. Rev. A\/} {\bf 82} 042332

\bibitem{Abramowitz1970}
Abramowitz M and Stegun I~A 1970 {Handbook of Mathematical Functions: With
  Formulas, Graphs, and Mathematical Tables Applied mathematics series} {\em
  Appl. Math. Ser.\/} vol~55 (United States Department of Commerce, National
  Bureau of Standards)

\bibitem{Forn-Diaz2016}
Forn-D{\'{i}}az P, Garc{\'{i}}a-Ripoll J~J, Peropadre B, Orgiazzi J~L, Yurtalan
  M~A, Belyansky R, Wilson C~M and Lupascu A 2017 {\em Nat. Phys.\/} {\bf 13}
  39--43 ISSN 1745-2473

\bibitem{Magazzu2018}
Magazz{\`{u}} L, Forn-D{\'{i}}az P, Belyansky R, Orgiazzi J~L, Yurtalan M~A,
  Otto M~R, Lupascu A, Wilson C~M and Grifoni M 2018 {\em Nat. Commun.\/} {\bf
  9} 1403 ISSN 2041-1723

\bibitem{PuertasMartinez2019}
{Puertas Mart{\'{i}}nez} J, L{\'{e}}ger S, Gheeraert N, Dassonneville R, Planat
  L, Foroughi F, Krupko Y, Buisson O, Naud C, Hasch-Guichard W, Florens S,
  Snyman I and Roch N 2019 {\em npj Quantum Inf.\/} {\bf 5}

\bibitem{FriskKockum2019}
{Frisk Kockum} A, Miranowicz A, {De Liberato} S, Savasta S and Nori F 2019 {\em
  Nat. Rev. Phys.\/} {\bf 1} 19--40

\bibitem{Forn-Diaz2019}
Forn-D{\'{i}}az P, Lamata L, Rico E, Kono J and Solano E 2019 {\em Rev. Mod.
  Phys.\/} {\bf 91} 025005

\bibitem{Viennot2018}
Viennot J~J, Ma X and Lehnert K~W 2018 {\em Phys. Rev. Lett.\/} {\bf 121}
  183601

\bibitem{Gaidarzhy2007}
Gaidarzhy A, Imboden M, Mohanty P, Rankin J and Sheldon B~W 2007 {\em Appl.
  Phys. Lett.\/} {\bf 91} 203503 ISSN 0003-6951

\bibitem{Rodriguez-Madrid2012}
Rodriguez-Madrid J~G, Iriarte G~F, Pedros J, Williams O~A, Brink D and Calle F
  2012 {\em IEEE Electron Device Lett.\/} {\bf 33} 495--497

\bibitem{Lo2018}
Lo L and Law C~K 2018 {\em Phys. Rev. A\/} {\bf 98} 063807 ISSN 2469-9926

\bibitem{Ferrari2019}
Ferreri A, Domina M, Rizzuto L and Passante R 2019 {\em Symmetry (Basel).\/}
  {\bf 11} 1384 ISSN 2073-8994

\bibitem{Souza2018}
Souza R~d~M~e, Impens F and Neto P~A~M 2018 {\em Phys. Rev. A\/} {\bf 97}
  032514 ISSN 2469-9926

  
  


\end{thebibliography}
\providecommand{\newblock}{}

\end{document}